\documentstyle[12pt,epsf]{article}
\def\lesssim{\mathrel{\mathpalette\vereq<}}
\def\vereq#1#2{\lower3pt\vbox{\baselineskip1.5pt \lineskip1.5pt
\ialign{$\hfill##\hfil$\crcr#2\crcr\sim\crcr}}}

\def\alt{\lesssim}

\begin{document}
\thispagestyle{empty}
\phantom{u}\hfill BI-TP 97/04\\
\phantom{u}\hfill JYFL 7/97\\
\phantom{u}\hfill nucl-th/9706012\\

\vspace*{1.7cm}
\centerline{ \bf INITIAL CONDITIONS IN THE ONE-FLUID HYDRODYNAMICAL}
\centerline{ \bf DESCRIPTION OF ULTRARELATIVISTIC NUCLEAR COLLISIONS}

\vspace*{1.0cm}

\centerline{Josef Sollfrank$^{a,}$\footnote{sollfran@physik.uni-bielefeld.de},
Pasi Huovinen$^{b,}$\footnote{huovinen@jyfl.jyu.fi} and 
P.V.~Ruuskanen$^{b,}$\footnote{ruuskanen@jyfl.jyu.fi} } 

\vspace*{0.6cm}

\centerline{\it 
$^a$ Department of Physics, University of Bielefeld, 
D-33615 Bielefeld, Germany}

\vspace*{0.05cm}

\centerline{\it 
$^b$ Department of Physics, University of Jyv\"askyl\"a
FIN--40351 Jyv\"askyl\"a, Finland}

\vspace*{1.2cm}
\noindent
\abstract{\noindent
We present a phenomenological model for the initial conditions
needed in a one-fluid hydrodynamical description of
ultrarelativistic nuclear collisions at CERN--SPS. The basic
ingredient is the parametrization of the baryon stopping, i.e.
the rapidity distribution, as a function of the thickness of
the nuclei. We apply the model to S + S
and Pb + Pb collisions and find after hydrodynamical evolution
reasonable agreement with the data.}

\vfill \noindent
Talk given at the {\it International Workshop on Applicability of 
Relativistic Hydrodynamical Models in Heavy Ion Physics}, 
ECT*, Trento (Italy), May 12 -- May 16, 1997

\clearpage
\newcommand{\an}{\langle n \rangle}

%%%%%%%%%%%%%%%%%%%%%%%%%%%%%%%%%%%%%%%%%%%%%%%%%%%%%%%%%%%%%%%%%
\section{Introduction}
%%%%%%%%%%%%%%%%%%%%%%%%%%%%%%%%%%%%%%%%%%%%%%%%%%%%%%%%%%%%%%%%%
One of the main motivations of heavy ion experiments is
the investigation of the nuclear equation of state away from
the ground state. However, it is very difficult to deduce from
the final multi-particle state, i.e. from the experimental
particle distributions, model independent statements about the
equation of state. It is not even granted, that the
time and volume available are large enough to create a form of matter
which can be described using (infinite volume) thermodynamics.
Nevertheless, the only possibility to obtain information
on the equation of state at large temperatures and densities
is via dynamical models of these collisions.

One popular approach to the dynamics of heavy ion collisions is
the use of one-fluid hydrodynamics \cite{Stoecker86,Schlei96,Sollfrank97a}.
Hydrodynamical simulations at energies below 10 $A$ GeV
(BEVALAC and SIS) usually start with the approaching nuclei
before their initial impact and include the initial compression.
In such a treatment the nuclei fuse to a single fluid,
implying, at zero impact parameter, a complete stopping of equal size
nuclei.  At higher energies, as at the CERN--SPS, RHIC and LHC, this is
not justified, and one must be able to incorporate nuclear
transparency in the description. Instead of trying to describe the
production and equilibration within hydrodynamics, one starts the
calculation from initial conditions which specify the hydrodynamic
state of the system at initial time $t_0$.  Initial conditions
parametrize the production and equilibration dynamics.

The aim of this article is to present a parametrization of the
initial baryon stopping, i.e. initial baryon rapidity distribution, in
terms of the thickness of the incoming nuclei.  The parametrization
should be valid for all symmetric collisions at a given cms energy.
Here we concentrate on CERN--SPS collisions at $\sqrt{s} \approx 20$
GeV per nucleon.  After having found such a parametrization there are
only a few additional parameters left like, e.g.~the initial time.
This makes the determination of the initial state from experimental
data, usually done by trial and error, less arbitrary.

%%%%%%%%%%%%%%%%%%%%%%%%%%%%%%%%%%%%%%%%%%%%%%%%%%%%%%%%%%%%%%
\section{Baryon Stopping}
%%%%%%%%%%%%%%%%%%%%%%%%%%%%%%%%%%%%%%%%%%%%%%%%%%%%%%%%%%%%%%
The experimental net baryon rapidity distribution $dN^{\rm B}/dy$ is
in general a function of $\sqrt{s}$ and the mass numbers $A$ and $B$
of the colliding nuclei.  The mass numbers determine the mean number
of interactions which an individual nucleon suffers.  The rapidity
distribution of final free baryons, $dN^{\rm B}/dy \, (\sqrt{s},A,B)$,
is measured in several experiments.  However, since we want to study
the evolution of the colliding system after its formation at time
$t_0$, we need the knowledge of $dN^{\rm B}/dy$ at $t_0$.  This
information can only be obtained by models, which describe the initial
production stage of the nuclear collision well.  Cascade type models
may in principle provide this information.  Our approach is much
simpler.  All physics of baryon stopping is included in the
parametrization of $dN^{\rm B}/dy$ with the final justification coming
from the comparison to the experiment.

\begin{figure}[tb]
 \vspace*{-1.0cm}
   \begin{center}\leavevmode
   \epsfysize=7.0cm \epsfbox{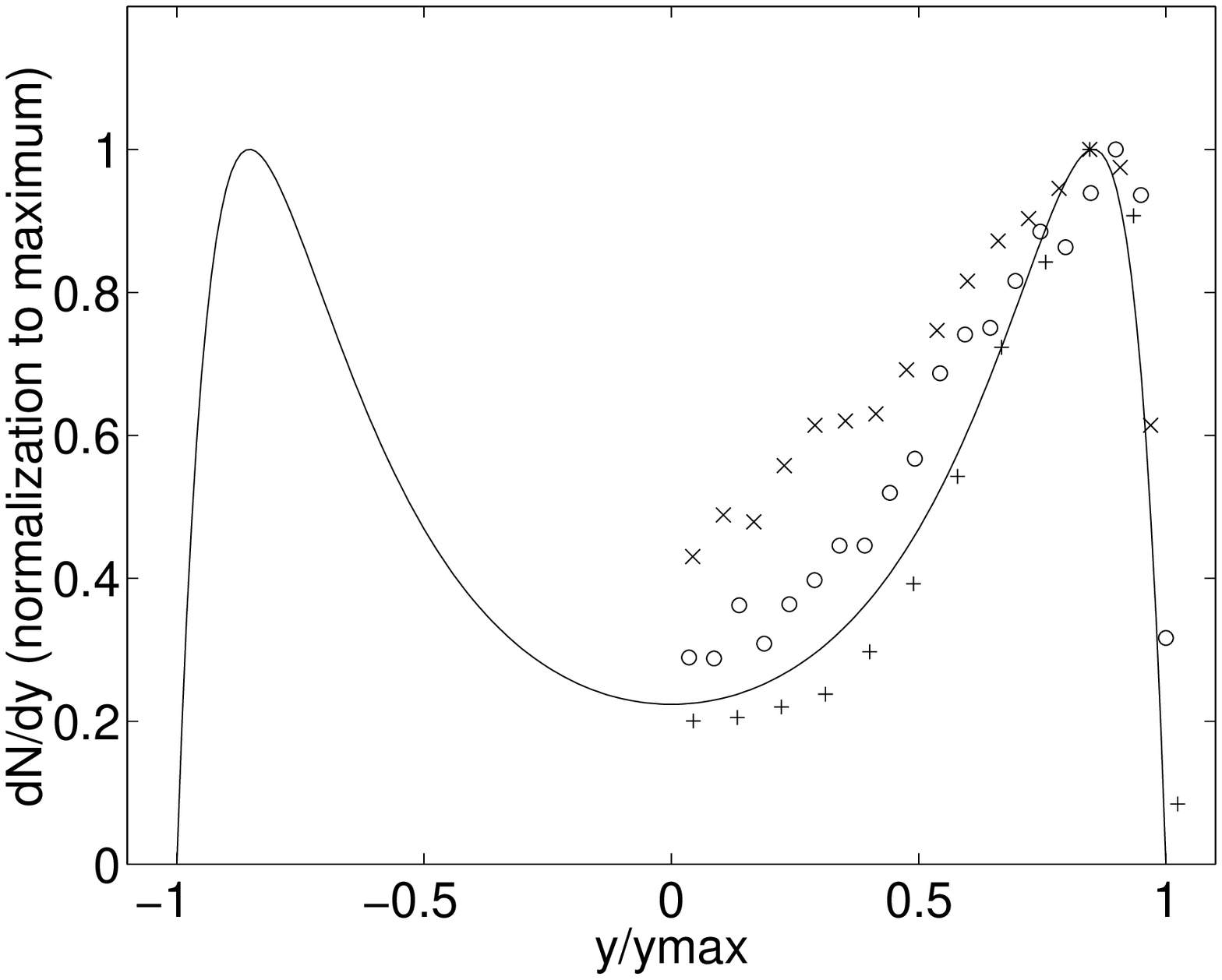} \hspace*{0.5cm}
   \epsfysize=7.0cm \epsfbox{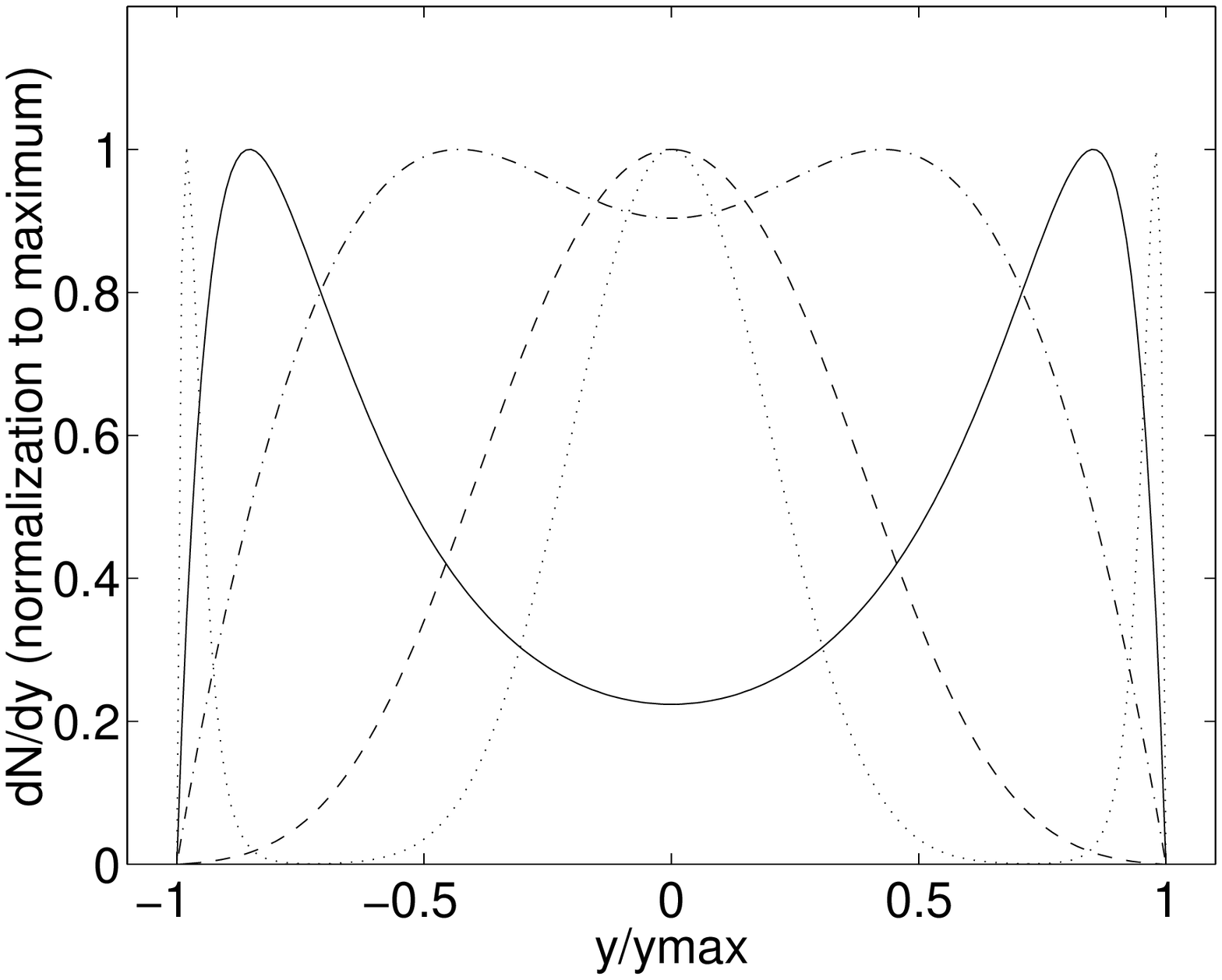}
   \end{center}

\vspace*{-1.9cm}
\begin{center}
\begin{minipage}{13cm}
\caption[]{\footnotesize \label{ppdata}
   a) Rapidity distribution of protons in $p + p$ collisions
   for three different energies normalized to their maximum:
   Experimental data: 'x' $\sqrt{s} = 4.9$ GeV \cite{Blobel74},
   'o' $\sqrt{s} = 6.9$ GeV \cite{Blobel74},
   '+' $\sqrt{s} = 27.5$ GeV \cite{Aguilar-Benitez91}; The line
   corresponds to the parametrization (\ref{param}) described in the text.\\
   b)
   Parametrization of the rapidity distribution of the initial
   baryons for various numbers of average nucleon--nucleon
   collisions $\an$ normalized to the maximum. Going from the edge to
   the center the lines correspond to $\an = 0.1$ (dotted), $\an =1$
   (full), $\an =2$ (dashed-dotted), $\an = 4$ (dashed) and
    $\an = 8$ (dotted). The line for $\an =1$
    is the same as the parametrization in Fig.~\ref{ppdata}a.
}
\end{minipage}
\end{center}
\end{figure}

To motivate our parametrization we start from what is known in
$p + p$ collisions. In order to compare various collisions at
different energies, we use the scaled rapidity $x_y = y/y_{max}$,
where $y$ is measured in the cms frame of the colliding protons.
In Fig.~\ref{ppdata}a we show the rapidity distribution
$dN/dy$ of protons for three values of $\sqrt{s}$, plotted against $x_y$ and
normalized to 1 at the maximum. Going from lower to higher energies,
the central region ($y\approx 0$) becomes more depleted indicating
increasing transparency of colliding nucleons with increasing $\sqrt s$.
The maximum, however, seems to stay  fixed around $x_y = 0.85$
independent of $\sqrt{s}$.

In Fig.~\ref{ppdata}a we also show a parametrization of the $dN/dy$ as
we interpolate the data to $\sqrt{s} = 20$ GeV as appropriate for
CERN--SPS nuclear collisions.  The following consideration went into
the parametrization. We compose the distribution out of two parts, one
dominant in the target and the other in the projectile fragmentation
region.  The rise in $dN/dy$ going from the center to forward (or
backward) regions is roughly exponential.  Therefore we start out with
two exponential functions of third order polynomials:
\begin{eqnarray}\label{param}
\frac{dN}{dy}
& = &
\frac{dN^{\rm Proj}}{dy} (x_y) + \frac{dN^{\rm Tar}}{dy} (x_y) = \\
&& \left [C_{\rm Proj} \exp(ax_y^3 + bx_y^2 + cx_y) +
C_{\rm Tar} \exp(-ax_y^3 + bx_y^2 - cx_y) \right] (1-x_y^2) \: . \nonumber
\end{eqnarray}
The factor $(1-x_y^2)$ is chosen to cut off the distribution at the
phase space boundary. The fit in Fig.~\ref{ppdata}a is given by
$a = 1.5$, $b = 0$, $c = 3.0$, and $C_{\rm Proj} = C_{\rm Tar}$ with
values of $C$'s determined via the normalization.

We now assume that the initial baryon rapidity distribution
$dN_{\rm B}/dy$ in a nuclear collision has the same functional
form as Eq.~(\ref{param}), but the parameters $a,b$ and $c$ depend
on the nuclear thickness a participating beam (target) nucleon sees
on its way through the target (beam) nucleus. The task is to find
this dependence.

The thickness function is defined as
\begin{equation}\label{thickfunction}
T_A (\vec{\rho}) = \int dz \; \rho_A(z, \vec{\rho}) \: ,
\end{equation}
where $\rho_A (\vec{r})$ is the nuclear density for a nucleus of
mass number $A$, $z$ is the longitudinal and $\vec\rho$ the transverse
variable: $\vec r=(z,\vec\rho)$. We take a Woods-Saxon parametrization
for the nuclear density
\begin{equation}\label{ws}
\rho_A(\vec{r}) = \frac{\rho_0}{\exp[(|\vec{r}| - R_A)/a_R] + 1} \: ,
\end{equation}
with
\begin{equation}\label{radiusdef}
R_A = 1.12 \: {\rm fm} \times A^{1/3} - 0.86
\: {\rm fm} \times A^{-1/3} \,
\end{equation}
$a_R = 0.54$ fm, and $\rho_0 = 0.17$ fm$^{-3}$ \cite{Bohr69}.

We expect that the stopping depends not only on the thickness but also
on an interaction strength, like the inelastic cross section. Then
the natural quantity for the parametrization of stopping is
\begin{equation}\label{defn}
\an (\vec{\rho}) = \sigma_{pp} \, T_{A}(\vec{\rho}) \; ,
\end{equation}
where we use $\sigma_{pp} = 32$ mb, the total inelastic cross section
for $p + p$ collisions at SPS energy.  In the Glauber model
\cite{Glauber59} $\an(\vec{\rho})$ is the average number of
interactions suffered by a nucleon colliding with a nucleus at impact
parameter $\vec b=\vec\rho$. The value of $\sigma_{pp}$ is not
important since $\an$ will be multiplied by adjustable parameters.
We should like to emphasize that $\an$ should be considered as a
reasonable quantity to describe the strength of stopping of a nucleon
and it need not be interpreted as the number of interactions except
at the limit of small $\an$. In particular, for
$\an = 1$ we should  have the same rapidity distribution as in
$p + p$ collisions.

We impose the following limits.  Since $|x_y| \le 1$ the parameter $a$
becomes important at the edge of the phase space.  For $T_{A}
\rightarrow \infty$ we expect $a$ to vanish because the nucleons are
shifted gradually to $x_y = 0$.  The limit $a \rightarrow 0$ is
achieved by $a \propto 1/T_{A}$.  More generally we could assume power
behaviour but the inverse dependence turns out to be sufficient.

In Eq.~(\ref{param}) the actual stopping is expressed through the
parameter
$b$ which is zero for $\an = 1$ and negative for $\an > 1$, and
should dominate for growing $\an$.  We use a simple ansatz $b =
{\beta_s}\: (1 - \an) = {\beta_s}\: (1 - \sigma_{pp} \, T_{A})$.  The
prefactor $\beta_s$ is a fit parameter characterizing the stopping.
From fits to the experimental particle spectra we obtain $\beta_s =
2.25$.  One has to keep in mind that the deduced value of $\beta_s$
depends on the time evolution from initial to final distributions
and therefore on the equation of state which is used.
Finally, the parameter $c$ could also be
chosen to depend on the thickness.  However, it turns out that we can
obtain the needed rapidity shift as a function of nuclear thickness
from the $\an$ dependence of $b(\an)$ alone and leave $c$ as a
constant.

We summarize the functional dependence of the parameters
$a,b$ and $c$ on $T_A$:
\begin{eqnarray}
a(T_A) & = & 1.5 \: (\sigma_{pp} \, T_{A})^{-1} \nonumber \\
b(T_A) & = & {\beta_s}\: (1 - \sigma_{pp} \, T_{A}) \nonumber \\
c(T_A) & = & 3.0  \: .\label{abc}
\end{eqnarray}
These functions (\ref{abc}) together with (\ref{param}) form our
phenomenological ansatz with the values of the parameters 
determined mainly from the S + S data.  In Fig.~\ref{ppdata}b we show, with
$\beta_s = 2.25$ the resulting $dN_{\rm B}/dy$ for several values of
$\an$ normalized as in Fig~1a.
For $\an = 1$ the curve is the same as in Fig.~\ref{ppdata}a.
The dependence of $dN_{\rm B}/dy$ on $\an$ shows the following limits.
For $\an \rightarrow 0$ the parameter $a(\an)$ dominates and leads
to a sharp rise at $|x_y| \alt 1$ which is cut down by the phase space
factor at the boundaries.  Therefore $dN_{\rm B}/dy$ turns to a double
$\delta$-function peaked at target and projectile rapidity.  For $\an
\rightarrow \infty$, the parameter $b(\an)$ dominates and results in a
Gaussian function for $dN_{\rm B}/dy$ which narrows with increasing
$\an$.

Since we obtain the distribution as a function of transverse variable
$dN_{\rm B}(\vec\rho)/dy$ it will be normalized to the number of
nucleons per unit transverse area at $\vec\rho$.
To obtain the final local density
we must know the dependence between rapidity and longitudinal distance
$z$ at the initial time $t_0$.  We do this by specifying the initial
velocity profile by choosing a linear relation between rapidity and
$z$:
\begin{equation}\label{vap}
y(\rho, z) = \kappa(\rho) z \,,\qquad
v_z(\rho,z) = \tanh[\kappa(\rho) z]
\:. \end{equation}
The local density $\rho_B(z,\rho)$ is obtained from the
$dN_{\rm B}(\vec\rho)/dy$ by multiplying with $(dy/dz)/\cosh y$
where $\cosh y$ is the Lorentz contraction factor between the local
rest frame and the fixed cms frame. We will specify the initial
flow velocity, in particular  the $\vec\rho$
dependence of $\kappa(\rho)$, after discussing the initial energy
distribution.

%%%%%%%%%%%%%%%%%%%%%%%%%%%%%%%%%%%%%%%%%%%%%%%%%%%%%%%%%%%%%%
\section{The Initial Energy Flow and Velocity Profile}\label{secinitial}
%%%%%%%%%%%%%%%%%%%%%%%%%%%%%%%%%%%%%%%%%%%%%%%%%%%%%%%%%%%%%%

The numerical solution of the hydrodynamical equations is determined
on a 2~+~1 dimensional grid in the cylindrical coordinate system
$\vec r= (z,\vec\rho)$ \cite{Sollfrank97a}. The frame which we use is
always the cms frame of the participating nucleons.

Since we assume azimuthal symmetry the simulation is strictly valid
only for zero impact parameter collisions.  However, even the
triggering to 5\% of most central events in the experiments corresponds to a
considerable range in impact parameter.  It turns out that if we
incorporate the full energy available in the impact parameter zero
collisions we slightly overshoot the experimental spectra.  To account
for the experimental impact parameter averaging, we use effective
nuclear sizes, i.e. we replace the incoming nuclei $A,B$ by $A^{\rm
eff} = \xi A$ and $B^{\rm eff} = \xi B$  and fix the geometry in
terms of effective mass numbers $A^{\rm eff}$ and $B^{\rm eff}$.  The
value of $\xi$ is close to one.  Our numerical algorithm for the
calculation of final particle spectra leads to a few per cent loss of
baryon number and energy \cite{Sollfrank97a}.  We compensate also for
these losses when fixing the value of $\xi$ to be 0.9 in symmetric
collisions.

For the hydrodynamical calculation we need also
the local energy  density $\varepsilon(\vec{r})$ and the initial
flow velocity at the initial time $t_0$.
For the energy density we follow a similar approach as for the
baryon density.  We first parametrize the energy per unit transverse
area and unit rapidity using a Gaussian distribution in rapidity
\begin{equation}\label{enparameter}
\frac{dE(\vec\rho)}{dy} = C_{\varepsilon} \exp\left[
\frac{-(y-y_0)^2}{2\sigma_\varepsilon^2} \right] \:
\left[1 - (y/y_{max})^2 \right] \: ,
\end{equation}
where the width $\sigma_\varepsilon$ and the normalization
$C_{\varepsilon}$ depend on the transverse coordinate $\vec\rho$.
The last factor is the same phase space cut-off as imposed for
the baryons (\ref{param}). Eq.~(\ref{enparameter}) contains only the
non-baryonic energy to which the contribution connected with the net
baryon density will
be added. The parametrization has three unknowns (at each $\vec\rho$),
but only one, e.g.~the
width $\sigma_\varepsilon$ may be freely chosen.  The normalization
$C_\varepsilon$ is given by energy conservation, imposed per unit
transverse area at each value of $\rho$. The value $y_0$ is
the center of mass rapidity for nucleons at $\vec\rho$.  For symmetric
collision $y_0 = 0$ in  the cms frame.

It turns out that the experimental pion rapidity distribution is
well reproduced when the value of
$\sigma_\varepsilon$ correlates with the
baryon stopping.  A larger stopping of baryon number results in a
larger stopping of energy, too.  Therefore the width of the energy
distribution in rapidity space decreases with increasing nuclear
stopping. We describe this with an ansatz
\begin{equation}\label{cy}
\sigma_\varepsilon(\rho) = \frac{c_\varepsilon}{
\left[\sigma_{pp}T_A(\rho)
\sigma_{pp}T_B(\rho)\right]^{\;\alpha_\varepsilon}}\:,
\end{equation}
where $\sigma_{pp}$ is included in order to make the denominator
dimensionless. The numbers appearing in (\ref{cy}) are determined from
the fits to the symmetric collision systems: $c_\varepsilon = 0.75$,
$\alpha_\varepsilon = 0.13$.

Next we add to the energy density Eq.~(\ref{cy}) the energy density
connected with the net baryon number.  This additional energy is
important at the edge of phase space and ensures that we take into
account the rest mass and the thermal energy of every leading nucleon.
We write the coordinate space density in the form
\begin{equation}\label{endef}
\varepsilon (\vec{\rho}, z) = \frac{dE(\rho)}{dy} \:
\frac{dy(\vec{\rho}, z)}{dz} + \zeta \rho_{\rm B} (\vec{\rho}, z) \: ,
\end{equation}
where $\zeta$ is the energy carried by the baryon. We take $\zeta=1.3
\ {\rm GeV}\approx m_{\rm nuc}+E_{\rm th}$
where $m_{\rm nuc}$ is the rest mass of a nucleon and $E_{\rm th}$
its thermal energy. Since $\rho_{\rm B}$
is the net baryon distribution, the energy of baryon-antibaryon pairs
is included in the first term.

The expansion of the matter is very sensitive on the
initial velocity profile. Since we consider zero impact parameter
collisions with cylindrical symmetry only, we do not expect significant
collective motion in transverse direction initially,
and take the velocity in the transverse
direction at $t_0$ to vanish, i.e., $\vec v_\rho(t_0,\vec{r})=0$.

In the case of the Bjorken model \cite{Bjorken83}, the scaling
ansatz for longitudinal velocity is $v_z = z/t$ and initial
conditions are usually defined on fixed proper time $\tau_0$.
At finite, albeit high collision energies, the longitudinal extent of
the system is finite and the scaling assumption must break
in the fragmentation regions.
Since we perform the numerical
calculations in the center-of-mass frame of the
participating nucleons we specify the initial condition at a fixed
time $t_0$ in this frame. On this line the scaling velocity is $z/t_0$.
Since scaling cannot hold, we have, for
convenience, chosen the linear $z$--dependence for rapidity instead
of velocity as given in Eq.~(\ref{vap}).
For small $z$ this ansatz can approach the velocity profile
of the scaling solution. The proportionality constant is now $\kappa$
instead of $1/t_0$ of the Bjorken model.
We define $\tau_0^{\rm eff} = 1/\kappa(\rho = 0)$ as
a parameter which can be regarded as an
equilibration time scale in the same way as $\tau_0$ in the scaling case.

\begin{figure}[tb]
%\vspace*{-1.0cm}
   \begin{center}\leavevmode
   \epsfysize=4.5cm \epsfbox{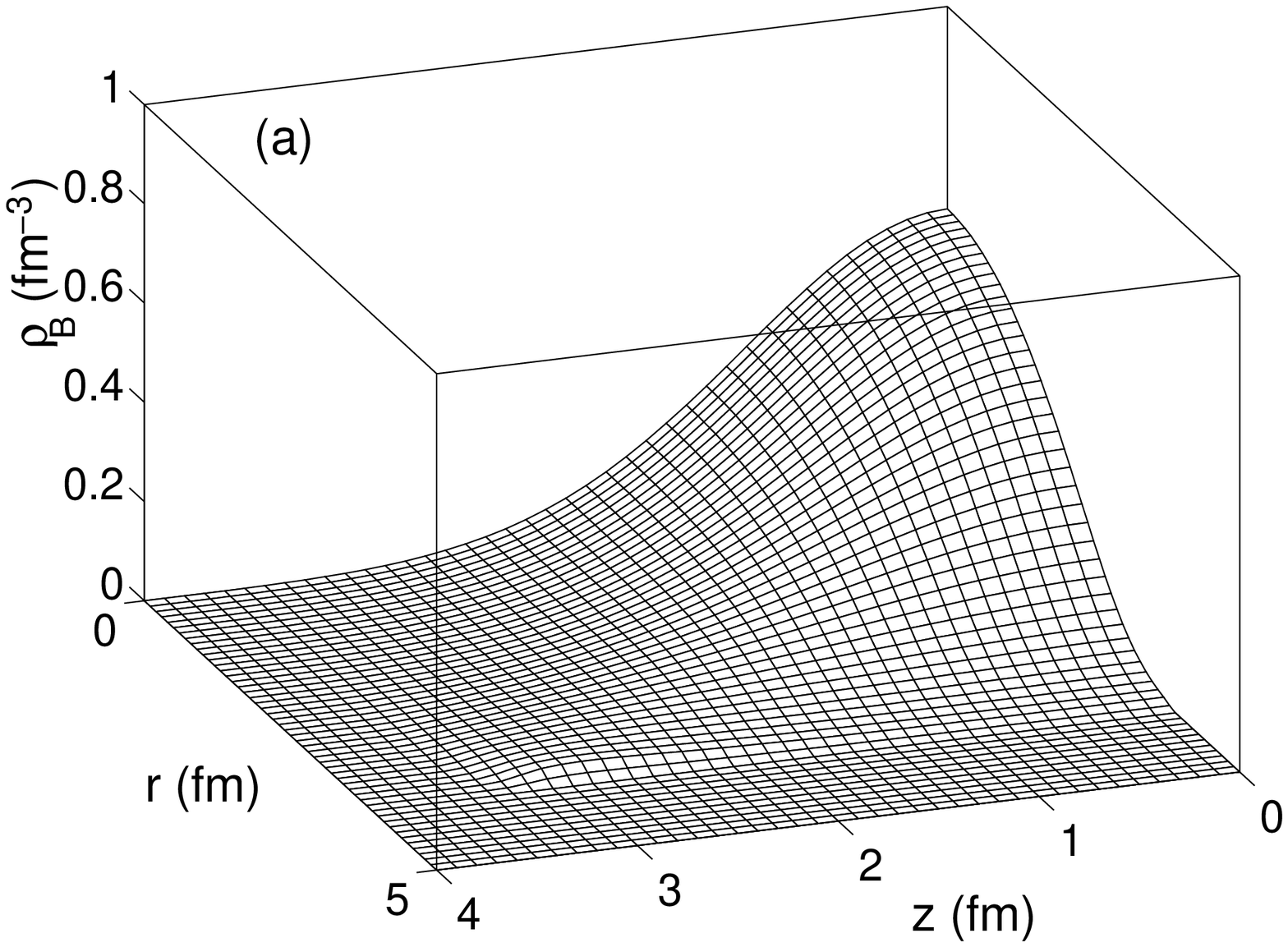}
   \epsfysize=4.5cm \epsfbox{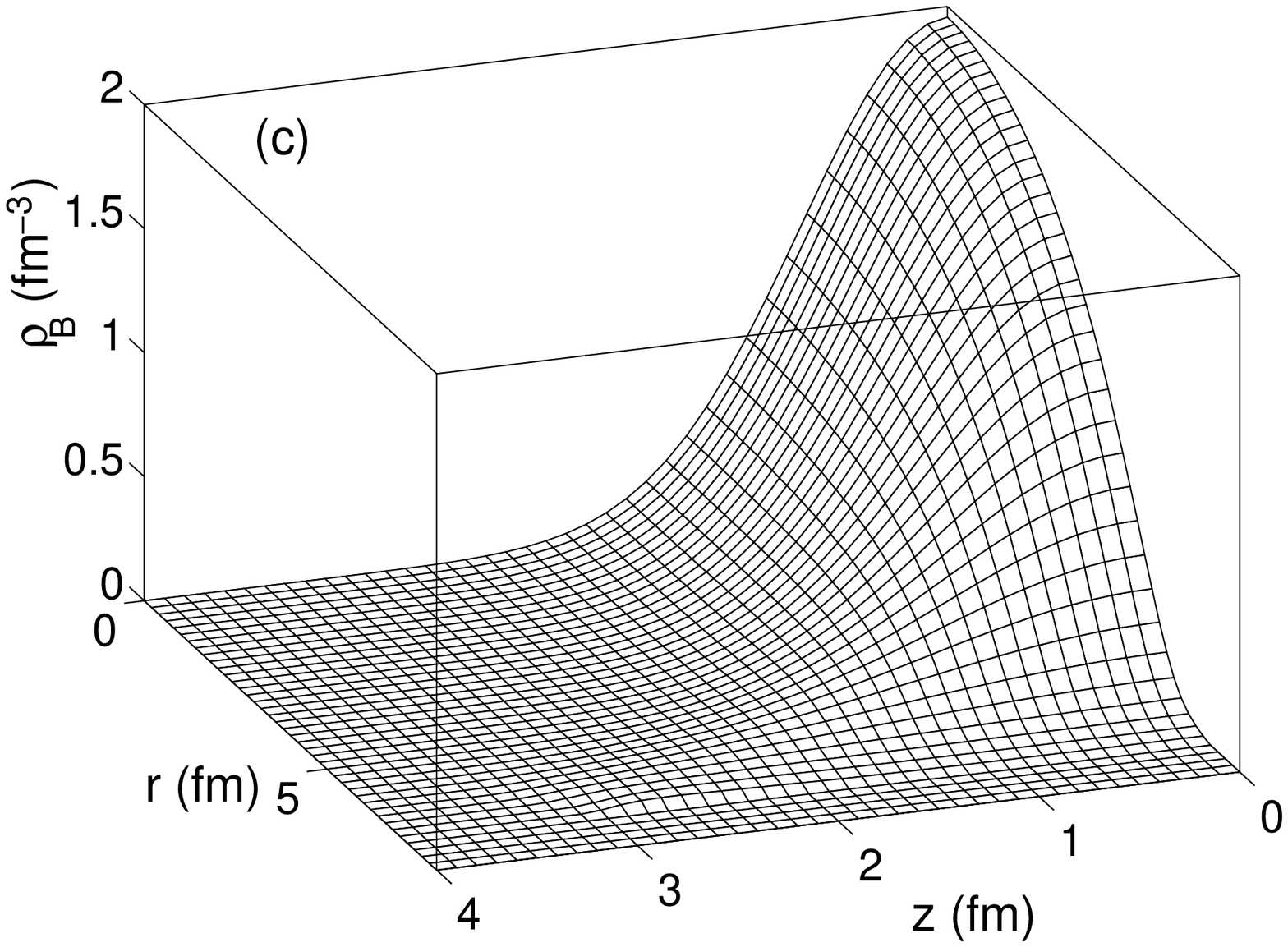}
   \end{center}

\vspace*{-0.5cm}
   \begin{center}\leavevmode
   \epsfysize=4.5cm \epsfbox{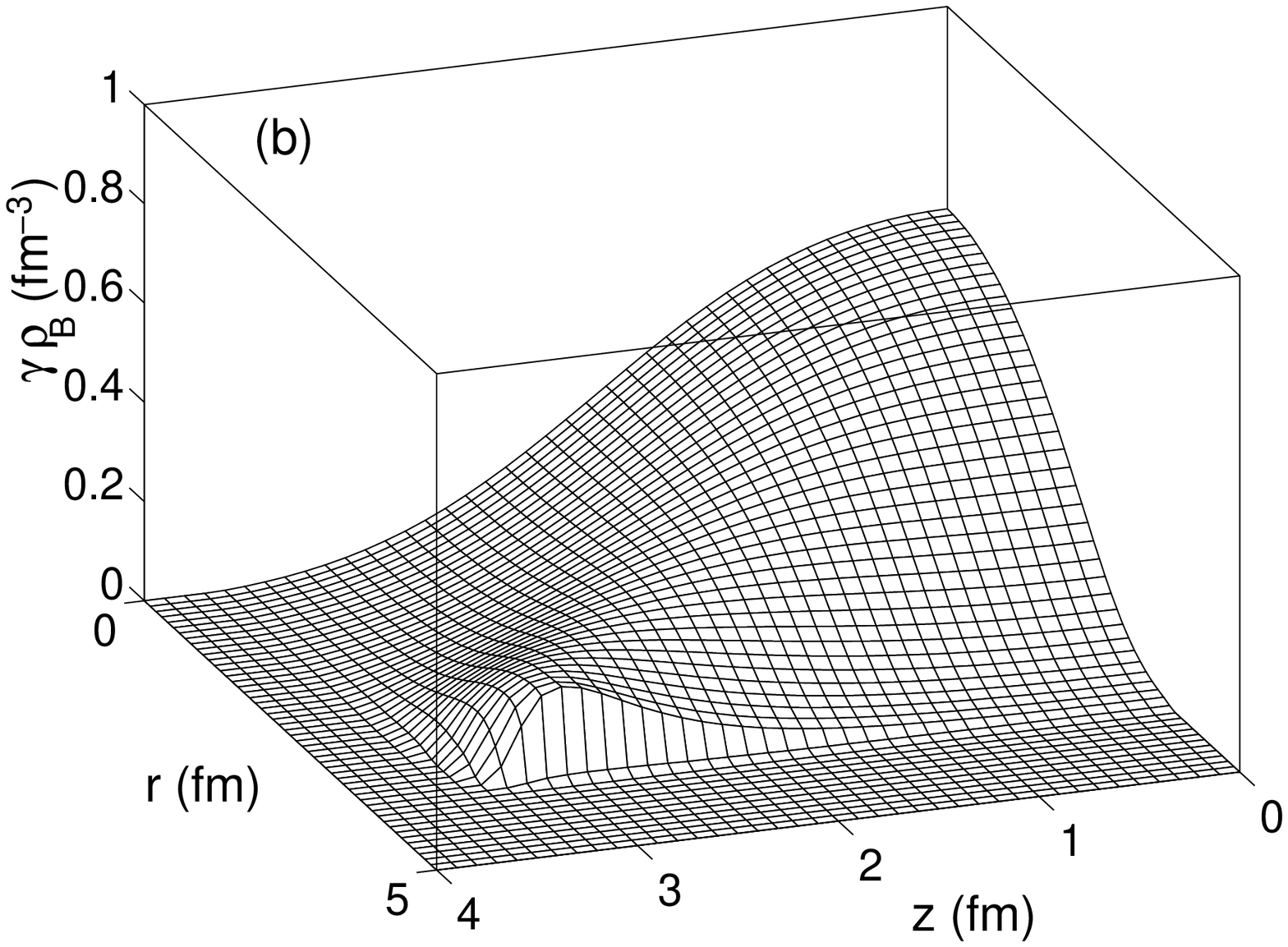}
   \epsfysize=4.5cm \epsfbox{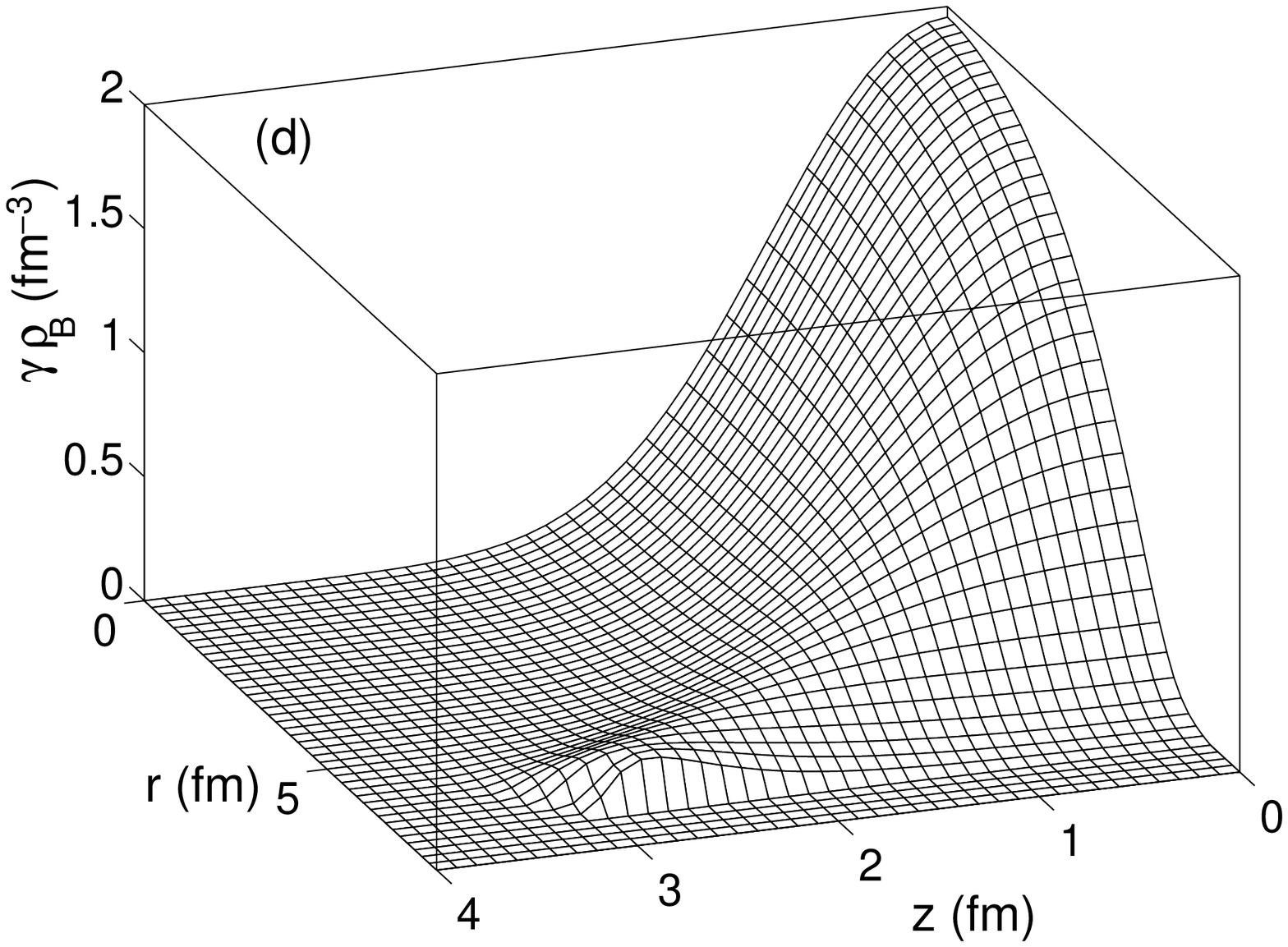}
   \end{center}

\vspace*{-1.0cm}
\begin{center}
\begin{minipage}{13cm}
\caption[]{\footnotesize \label{initialps}
  Local baryon density $\rho_{\rm B}$ (upper) and $\gamma \rho_{\rm
  B}$ (lower) in the $z$-$\rho$ plane as they result from the
  parametrization.
  The initial distribution is for S~+~S collisions (left) and
  Pb~+~Pb (right).
}
\end{minipage}
\end{center}
\end{figure}

We determine the $\rho$ dependence of $\kappa(\rho)$
through the longitudinal extension of the system, $z_L(\rho)$,
at different values of $\rho$. The $\rho$ dependence arises from the
variation of the nuclear thickness in the transverse plane.
The difference of the longitudinal extension between the center and
the edge is of the
order of $\Delta z =  z_L(\rho = 0) - z_L(\rho = R_A) \approx R_A/\gamma$,
where $\gamma = \cosh(y_{A})$ is the Lorentz contraction of the
nucleus in the fireball rest frame.
This form follows from nuclear geometry and the assumption that formation
and equilibration times are independent of $\rho$.
We define
\begin{equation}\label{defzl}
z_L(\rho) = z_0 + \frac{\sqrt{R_A^2 - \rho^2}}{\gamma} \: ,
\end{equation}
where $z_0$ is the longitudinal extension of the matter at equilibration
time in the thin disk limit. Instead of $z_0$ we use $\tau_0^{\rm eff}$
as an independent parameter with the  assumption that the forward (backward)
edge of the initial fireball, $z_L(\rho)$,
moves with the target (projectile) rapidity $y_{\rm cms}$, which,
from Eq.~(\ref{vap}) at $\rho=0$ gives
\begin{equation}\label{znull}
z_0 = \tau_0^{\rm eff} \; y_{\rm cms} - \frac{R_A}{\gamma} \: .
\end{equation}
Combining the above definitions one gets
\begin{equation}\label{defkappa}
\kappa(\rho) = \frac{y_{\rm cms}}{z_L(\rho)} \: .
\end{equation}

Even though  the initial distributions approach zero in transverse
direction as a consequence of the finite size of the nuclei
(cf.~Eq.~(\ref{ws})), it is still practical
for the numerical work to make them zero beyond some value of $\rho$.
As a criterium we use the mean value of interactions, $\an$ and
set the initial distributions to zero when
$\an < 0.5$. It is clear that even so the matter at the edge
is not dense enough for thermalization but for a finite system we
are forced to extend the calculation at the surface to densities were
the hydrodynamics cannot be justified.

We show in Fig.~\ref{initialps} the resulting initial baryon density
distributions for S~+~S and Pb~+~Pb.  The upper frames give the
densities in the local comoving frame and the lower the spatial
density in the overall rest frame of the collision.  The difference 
comes from the $\gamma$ factor of the Lorentz contracted fluid cell.
The figure clearly shows the increase in transparency with radius.  We
think this is more reasonable than using the same baryon distribution
at the edge as at the center.  In this figure the cut in $\an$ leads
to a sharp discontinuity in radial directions near the maximum.
However, this does not influence the dynamics, because
the pressure gradients are given by the local densities,
which are small and smooth in this region when going to larger $\rho$
as seen in the two upper figures.

In Tab.~\ref{table} we give the values of parameters which characterize
the initial state. An interesting point is that $\tau^{\rm eff}_0$
which is a parameter adjusted to each collision system turns
out to be rather similar in Pb~+~Pb and S~+~S collisions. This means
that the equilibration time seems to be independent of the thickness
of the nuclei.

\begin{table}[ht]
\begin{center}
\begin{minipage}{13cm}
\caption[]{\footnotesize \label{table}
Summary of parameters characterizing the collision when
a QGP equation of state is used. $\xi$ and $\tau^{\rm eff}_0$
are parameters adjusted to the data and explained in the text. The
rest follows from the initial parametrization. $t_f$ is the central
freeze-out time and $\langle v_\rho \rangle$
the mean radial velocity in the rapidity range of $|y_z|<0.25$.
}
\end{minipage}

\vspace*{0.3cm}
\begin{minipage}{.8\textwidth}
\renewcommand{\footnoterule}{\kern -3pt}
\begin{tabular}{lll}
\hline\\[-10pt]
collision  & S + S    &   Pb + Pb \\
\hline\\[-10pt]
$\xi      $  & 0.9       &      0.9  \\
$\tau^{\rm eff}_0$ (fm/c)&1.2&  1.3  \\
\hline
$z_L(0)$ (fm)
             &  3.5       &    3.6   \\
cent. $\varepsilon$ (GeV/fm$^3$)
             &  7.1      &   16.7   \\
cent. $\rho_{\rm B}$ (fm$^{-3}$)
             &  0.59     &   1.94   \\
cent. $T$ (MeV)
             &  244      &   300   \\
$R(\xi A)$ (fm)
             &  3.15     &   6.25   \\
\hline
$t_f$ (fm/c)
             &  6.1      &  11.4  \\
$\langle v_\rho \rangle (|y_z|<0.25)$
             &  0.28     &  0.34  \\
\hline
\end{tabular}
\end{minipage}
\end{center}
\end{table}

%%%%%%%%%%%%%%%%%%%%%%%%%%%%%%%%%%%%%%%%%%%%%%%%%%%%%%%%%%%%%%
\section{Results}
%%%%%%%%%%%%%%%%%%%%%%%%%%%%%%%%%%%%%%%%%%%%%%%%%%%%%%%%%%%%%%

\begin{figure}[tb]
%\vspace*{-1.0cm}
   \begin{center}\leavevmode
   \epsfysize=4.5cm \epsfbox{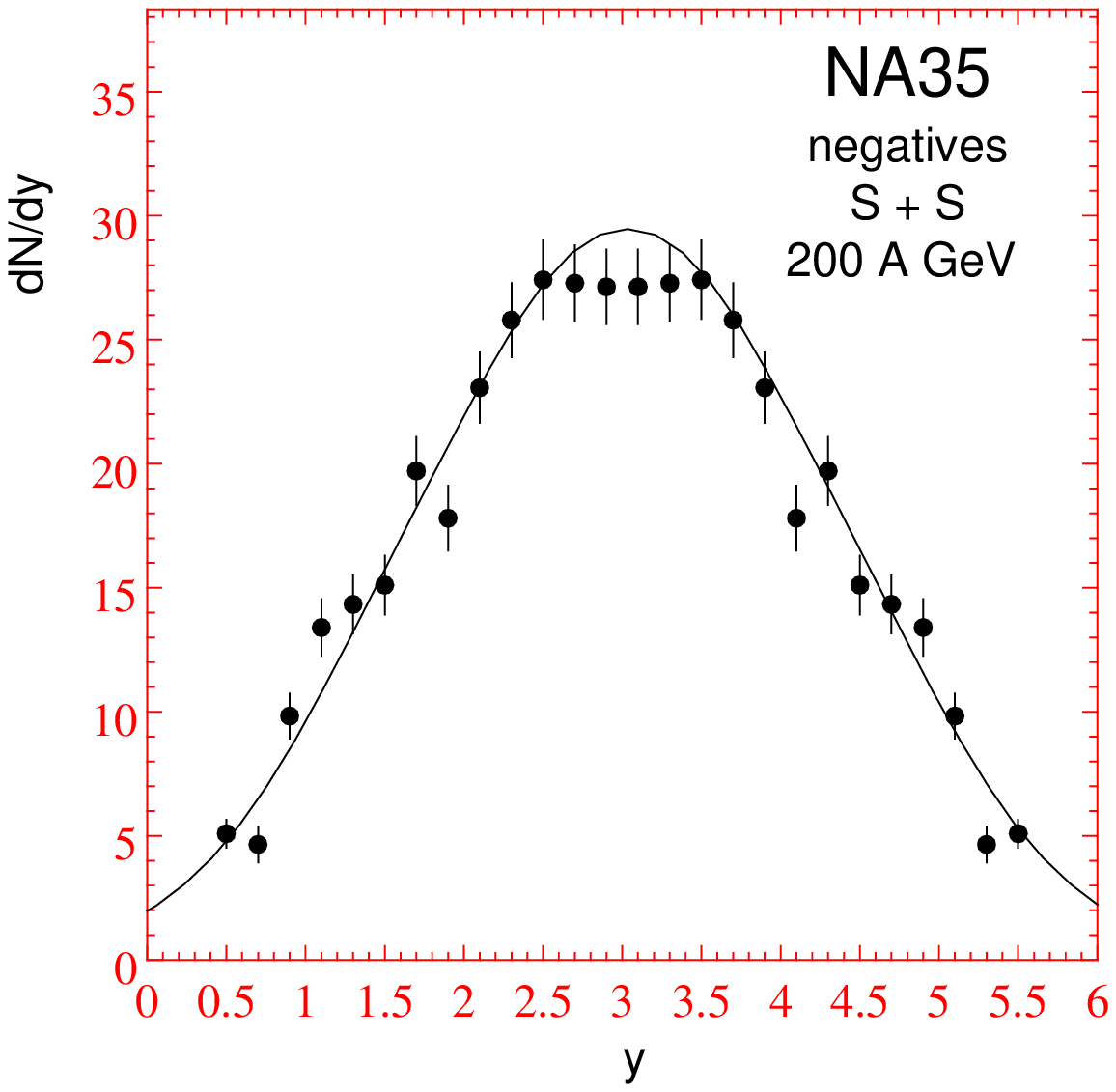}
   \epsfysize=4.5cm \epsfbox{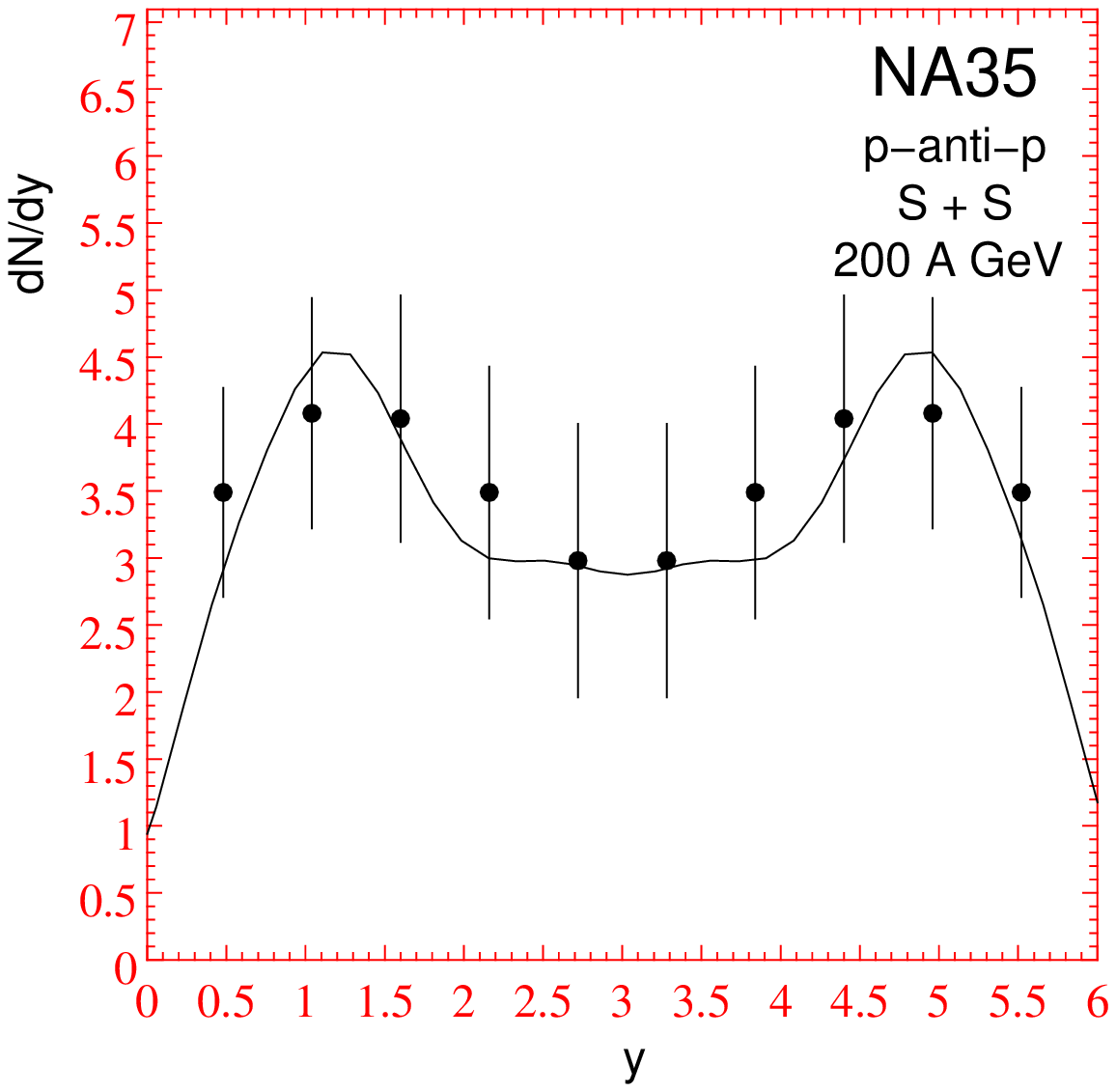}
   \end{center}

\vspace*{-0.5cm}
   \begin{center}\leavevmode
   \epsfysize=4.5cm \epsfbox{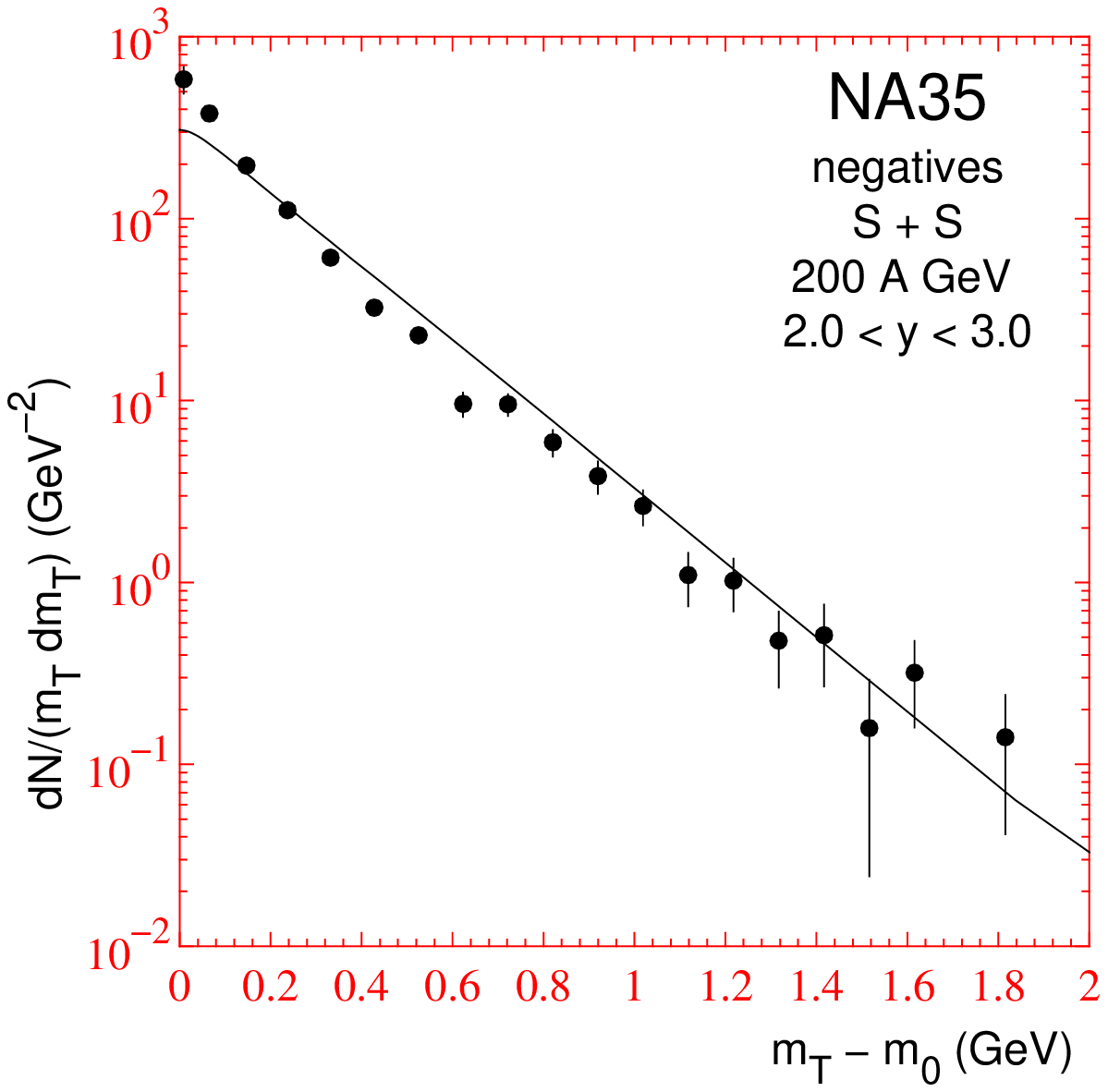}
   \epsfysize=4.5cm \epsfbox{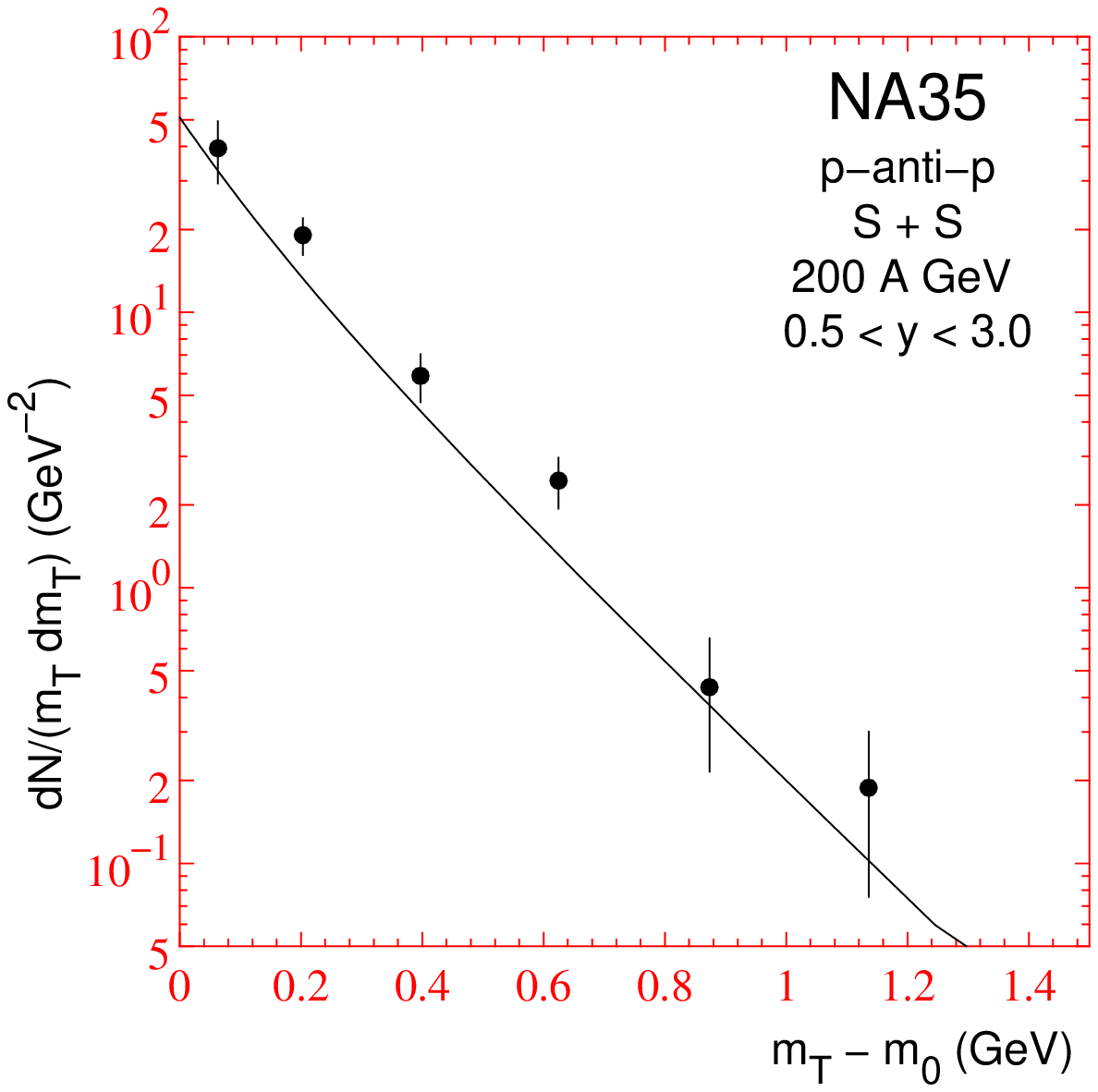}
   \end{center}

\vspace*{-1.0cm}
\begin{center}
\begin{minipage}{13cm}
\caption[]{\footnotesize \label{ss}
  Comparison of experimental data from S~+~S collisions with
  final hadron spectra  calculated using an equation of
  state (EOS A) with phase transition at $T_c=165$ MeV.
  The data are taken by the NA35 collaboration \cite{Bachler94}.
}
\end{minipage}
\end{center}
\end{figure}

We illustrate the use of the parametrization of initial conditions
discussed above by comparing with experimental data  from the NA35
\cite{Bachler94} and NA49 \cite{Jones96} collaborations.
We solve the hydrodynamical evolution as described in \cite{Sollfrank97a},
where also the treatment of the freeze-out is explained. We use
isospin symmetry when calculating the hadron spectra. The isospin
corrections would somewhat reduce the
calculated $p - \bar{p}$ spectrum in the heavy systems
(e.g. Pb + Pb) but our choice $\xi = 0.9$ probably
overestimates the amount of spectator nucleons.
The freeze-out happens at a constant energy
density of $\varepsilon_{f} = 0.15 \; {\rm GeV/fm}^3$ leading to an
average freeze-out temperature of $T_f \approx 140$ MeV. In
\cite{Sollfrank97a} we also explain the construction of the equation of
state we use. Here we show results only for an equation of state with phase
transition to QGP at $T_c=165$ MeV, labeled as EOS A in \cite{Sollfrank97a}.

The stopping shows up mainly in the
rapidity distributions. We show in Fig.~\ref{ss} the
distributions for negatives and  proton-antiproton difference
in a central S + S collision.
In the $p - \bar{p}$ rapidity distribution
target and projectile fragmentation regions exhibit clear maxima, which
are nicely reproduced by the calculation.
The normalization is slightly adjusted by the $\xi$ parameter.
The rapidity distribution of the negatives fits very well.
Also the transverse momentum distributions of both negatives
and $p-\bar p$ are reproduced reasonably well.

\begin{figure}[tb]
%\vspace*{-1.0cm}
   \begin{center}\leavevmode
   \epsfysize=4.5cm \epsfbox{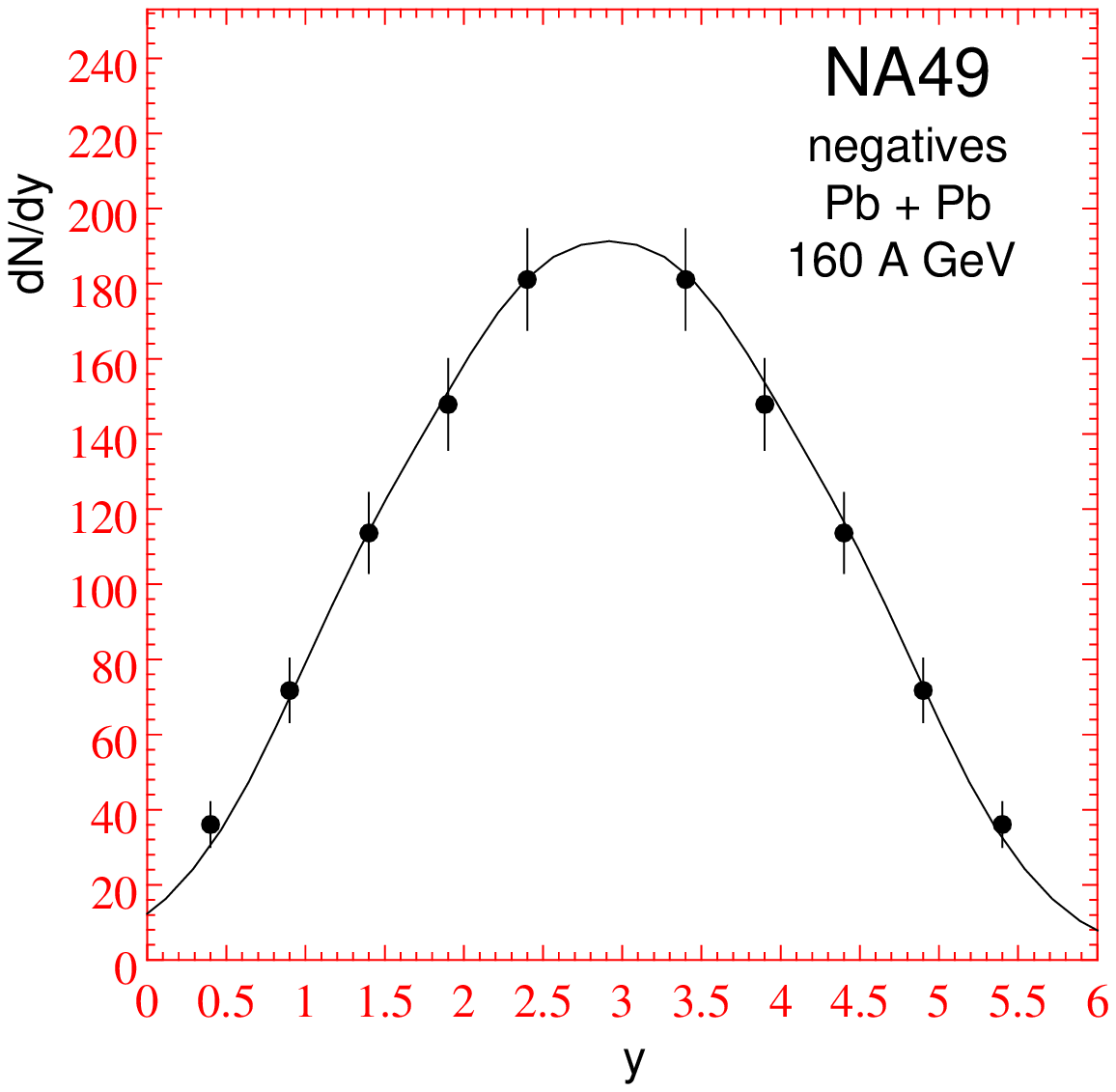}
   \epsfysize=4.5cm \epsfbox{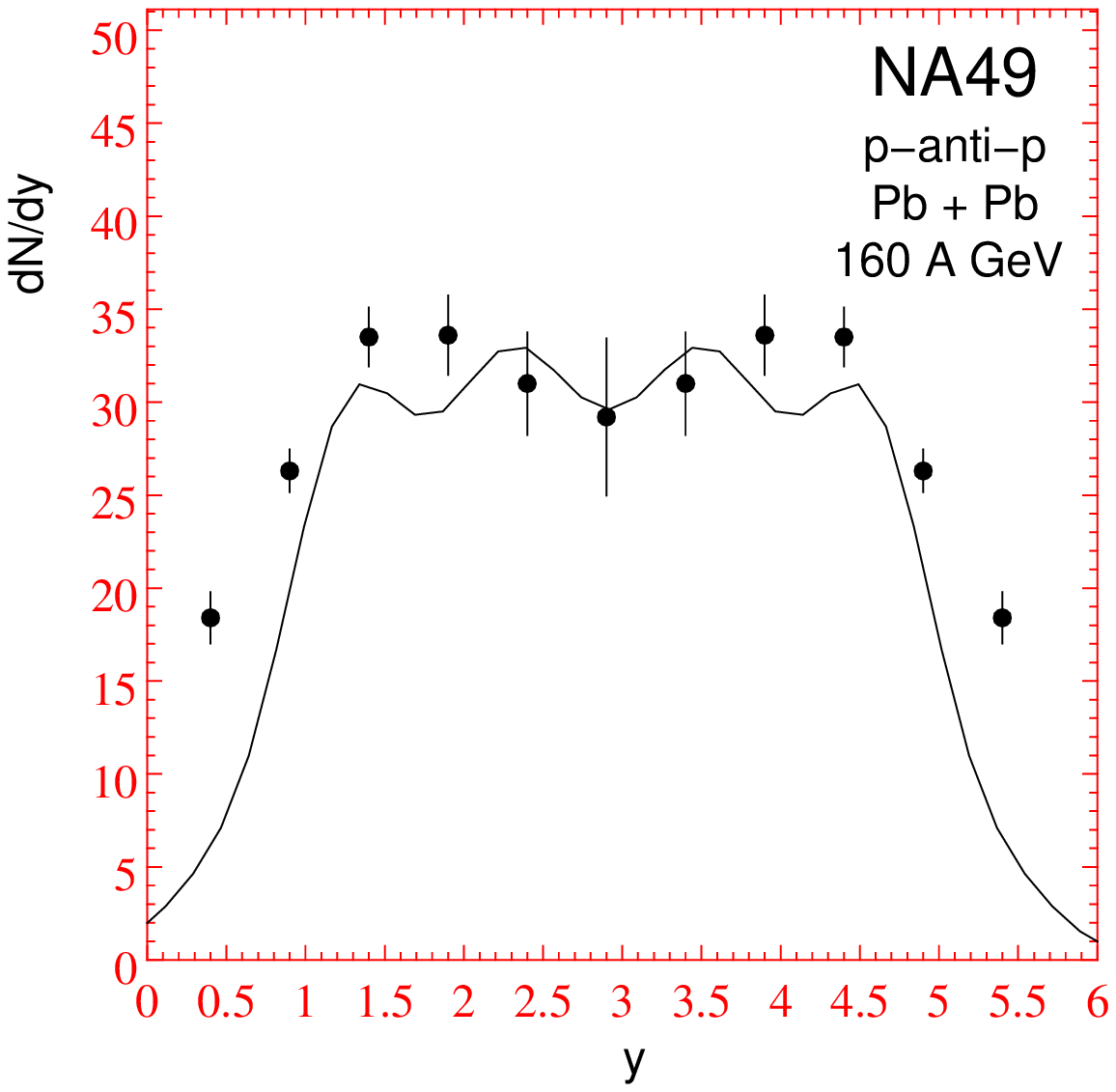}
   \end{center}

\vspace*{-0.5cm}
   \begin{center}\leavevmode
   \epsfysize=4.5cm \epsfbox{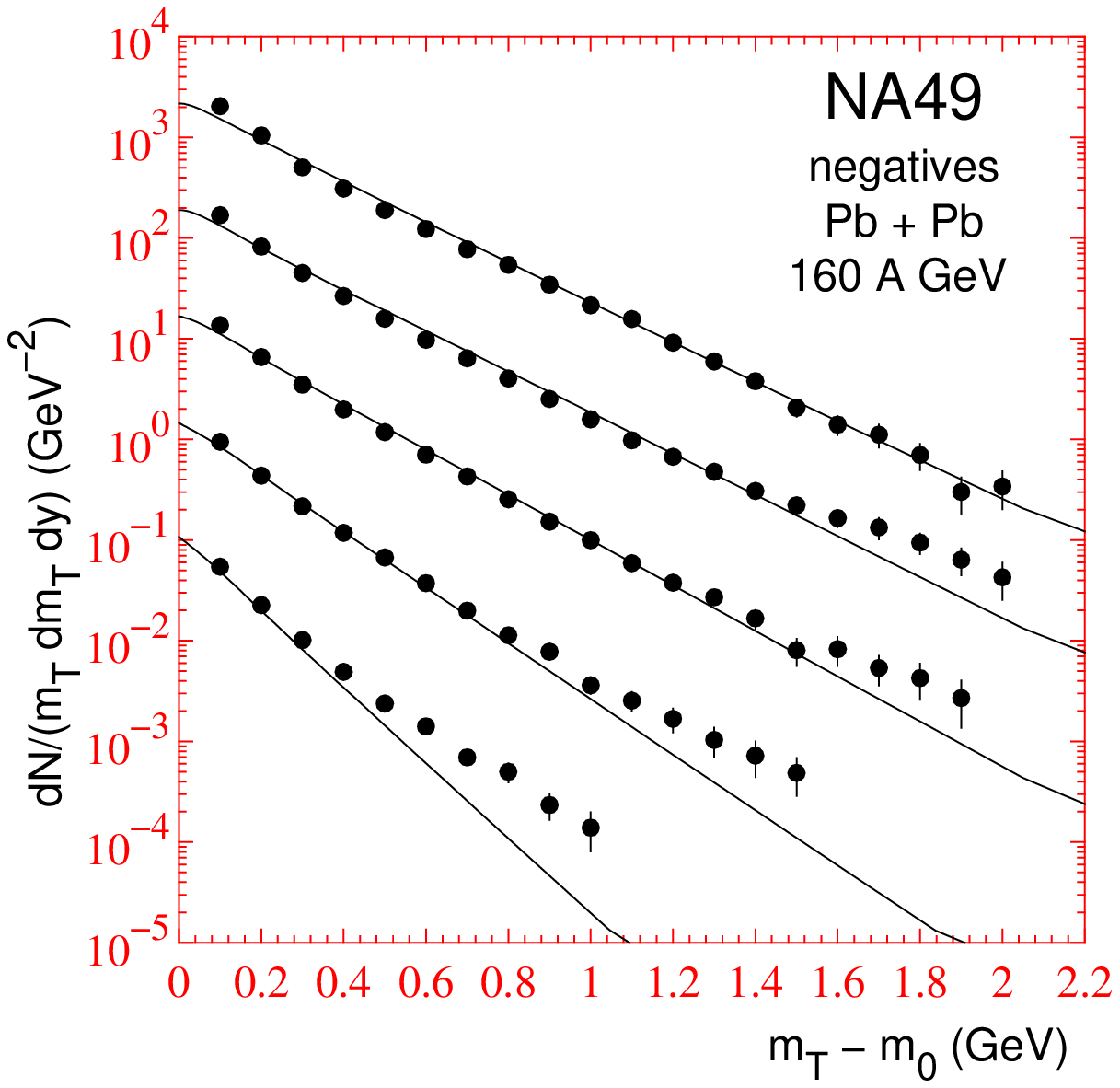}
   \epsfysize=4.5cm \epsfbox{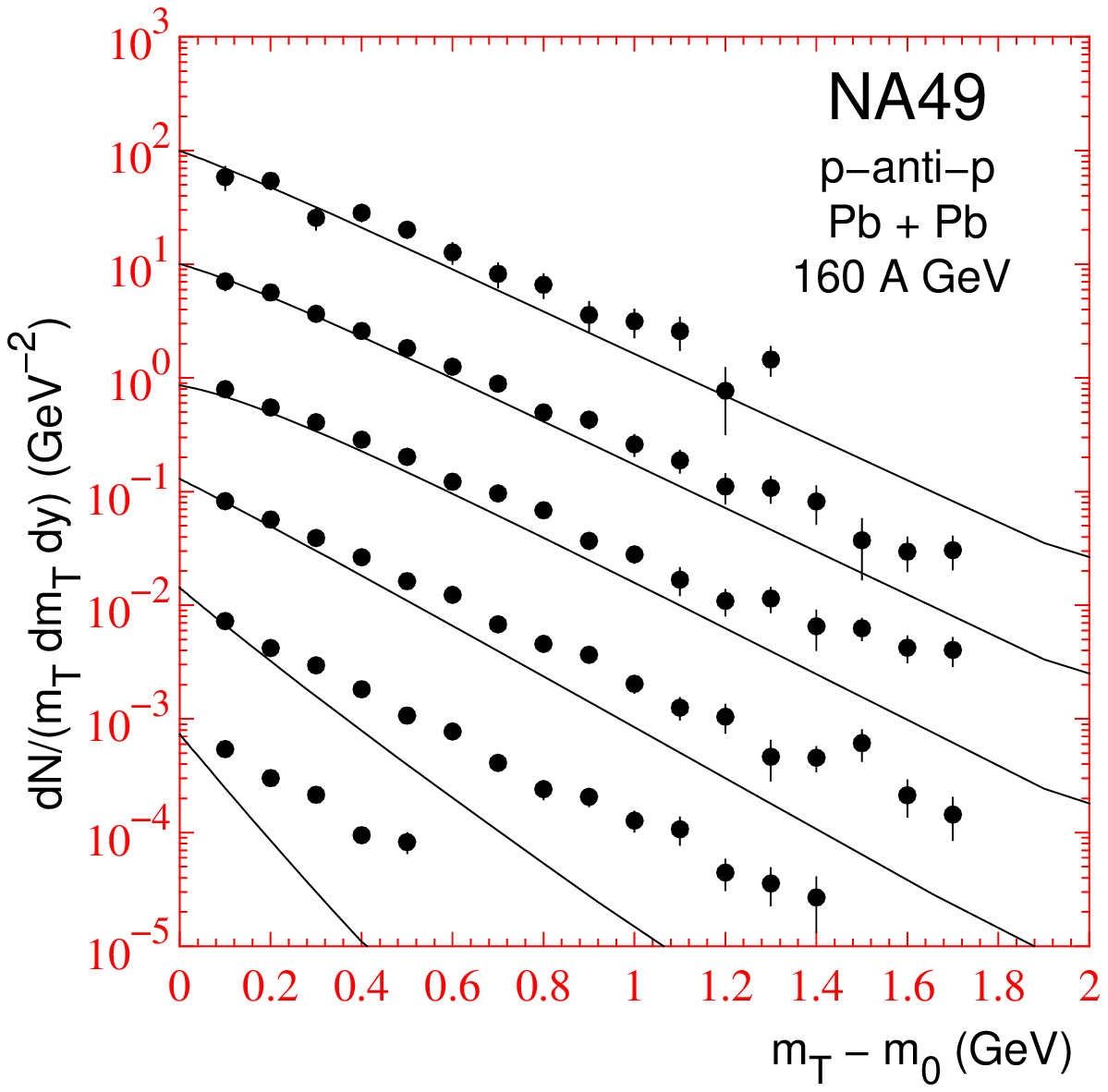}
   \end{center}

\vspace*{-1.0cm}
\begin{center}
\begin{minipage}{13cm}
\caption[]{\footnotesize \label{pbpb}
  Comparison of experimental data from Pb~=~Pb collisions with
  final hadron spectra  calculated using an equation of
  state (EOS A) with phase transition at $T_c=165$ MeV.
  The data are taken by the NA49 collaboration \cite{Jones96} and
  should be considered as preliminary. The
  $m_T$-spectra of negatives and $p - \bar{p}$ are measured
  in rapidity intervals of width 0.5
  centered at 2.9 ($p - \bar{p}$ only), 3.4, 3.9, 4.4, 4.9, 5.4.
  For clarity the data and the calculation are successively scaled
  down by $10^{-n}$ ($n = 0$,1,2,...).
}
\end{minipage}
\end{center}
\end{figure}

Fig.~\ref{pbpb} shows the result for Pb + Pb collisions compared
with preliminary data from the NA49 collaboration \cite{Jones96}.
There are small deviations in the proton distribution but
the negative particle rapidity distribution is very well reproduced.
{\it Both collisions are calculated with nearly the same parametrization},
the difference coming from the thickness function $T_A(\rho)$
only. The only slight difference is the $\tau_0^{\rm eff} = 1.2$ for
S~+~S and $\tau_0^{\rm eff} = 1.3$ for Pb~+~Pb.
Thus the parametrization reproduces well the stopping as a function
of nuclear size. The result for $p-\bar p$ show
larger deviation with faster protons in the experimental spectrum
than in the calculated result. At least part of this must come from the
finite impact parameter range included in the measurement and leading
to a larger average transparency. In the calculation we take this into
account only in the normalization through the parameter $\xi$ with
practically no change in the shape of the spectrum.

The gross features of the transverse momentum data of the
negative particles are again well reproduced except for
the very forward (and backward) region were
the calculated slope is steeper than the data.
The particle density at this edge of the phase space, however, is
becoming so small that the hydrodynamics with relatively strong longitudinal
flow can lead to an artificially large transverse cooling.
Thus we should not expect that hydrodynamics can describe the far edges
of fragmentation regions well.

This problem is more pronounced
in the $p - \bar{p}$ transverse distributions which indicate that even
in the central rapidity region the experimental transverse flow is somewhat
stronger. A possible explanations is that the equation of state is too soft.
Later freeze-out time might also enhance the flow effect on protons
as compared to pions.

In Tab.~\ref{table} we also give numbers which characterize the freeze-out.
The lifetime $t_f$ of the fireball, measured at the center, scales
as expected with the size of the fireball. We see a doubling of
the lifetime going from S + S to Pb + Pb in accordance with the doubled
radius of the lead nucleus. We further give the average (radial)
transverse velocity $\langle v_\rho \rangle$ of the fluid cells at
freeze-out with rapidity $|y_z| < 0.25$. The result may be
compared with an analysis of NA44 \cite{Bearden96}. They extracted
from a fit to hadron data a mean transverse velocity. The result is
\cite{Bearden96} $\langle v_\rho \rangle = 0.24 \pm 0.10 $ for S + S
and  $\langle v_\rho \rangle = 0.36 \pm 0.14$ for Pb + Pb in agreement
with our studies. However, the NA44 obtained a higher freeze-out
temperature for
Pb + Pb, $T_f = 167 \pm 13$ MeV (for S + S $T_f = 142 \pm 5$ MeV).
This explains why NA44 gets agreement with $\langle v_\rho \rangle = 0.36$
while we slightly underestimate the slope in the central $p-\bar p$
transverse momentum spectra.

Finally we compare our results with the related work of Schlei et
al.~\cite{Schlei96}.  (See also \cite{Schlei97} in this Volume, where
a different parameter set for initial conditions is used.)  The main
difference is in the geometry of the initial fireball.  The equations
of state differ slightly.  Both exhibit a phase transition to QGP,
ours with $T_c=165$ MeV vs.\ 200 MeV in~\cite{Schlei96}.  In addition,
the baryonic pressure in \cite{Schlei96} is omitted.

The basic differences may be summarized in the following way.  The
calculation in \cite{Schlei96} starts from a small longitudinal
extension, indicating large compression, and resulting in larger
initial energy density as compared to ours.  At the center of the
fireball for S+S collision $\epsilon_i = 13.0$ GeV/fm$^3$ in
\cite{Schlei96} vs.
7.1 GeV/fm$^3$ in our case.  For Pb+Pb the numbers are 20.4 GeV/fm$^3$
and 16.7 GeV/fm$^3$, respectively.  At first glance this looks like a
difference in stopping but a closer inspection shows that it is
related more closely to the initial volume than to the initial
velocity
distribution.  This can be interpreted as a difference in initial
time, our calculation corresponding to a later initial time.  If the
energy stopping is characterized with the fraction of thermal energy,
the numbers are very similar: 0.43 (0.64) in \cite{Schlei96} vs. our
0.45
(0.66) for S+S (Pb+Pb).  The difference in initial volume is related
to the longitudinal extent and our larger initial volume can be
interpreted as
starting the calculation at later initial time.  This shows up also
as shorter lifetimes in our calculation, 6.1 fm/c vs.\ 6.9 fm/c in
\cite{Schlei96} for S+S and 11.4fm/c vs.\ 13.5 fm/c in \cite{Schlei96}
for Pb+Pb collisions.

%%%%%%%%%%%%%%%%%%%%%%%%%%%%%%%%%%%%%%%%%%%%%%%%%%%%%%%%%%%%%%
\section{Conclusions}
%%%%%%%%%%%%%%%%%%%%%%%%%%%%%%%%%%%%%%%%%%%%%%%%%%%%%%%%%%%%%%

A model for initial conditions was developed with
a parametrization of baryon stopping in terms of the nuclear thickness
as the basic input. The {\it same} para\-me\-tri\-za\-tion gives a very 
satisfactory
description of the basic features of hadron spectra both in S~+~S
and Pb~+~Pb collisions. We think that this is an improvement compared
to earlier approaches \cite{Schlei96,Sollfrank97a} where the transverse
dependence of stopping was not taken into account.

\vspace*{0.5cm}\noindent{ACKNOWLEDGMENT}\\
This work was supported by the Bundesministerium f\"ur Bildung und
Forschung (BMBF) grand no. 06 BI 556 (6) and the Academy of Finland
grant no. 27574.

%%%%%%%%%%%%%%%%%%%%%%%%%%%%%%%%%%%%%%%%%%%%%%%%%%%%%%%%%%%%%%%%%%%%

\newcommand{\IJMPA}[3]{{ Int.~J.~Mod.~Phys.} {\bf A#1}, #3 (#2)}
\newcommand{\JPG}[3]{{ J.~Phys. G} {\bf {#1}}, #3 (#2)}
\newcommand{\AP}[3]{{ Ann.~Phys. (NY)} {\bf {#1}}, #3 (#2)}
\newcommand{\NPA}[3]{{ Nucl.~Phys.} {\bf A{#1}}, #3 (#2)}
\newcommand{\NPB}[3]{{ Nucl.~Phys.} {\bf B{#1}}, #3 (#2)}
\newcommand{\PLB}[3]{{ Phys.~Lett.} {\bf {#1}B}, #3 (#2)}
\newcommand{\PRv}[3]{{ Phys.~Rev.} {\bf {#1}}, #3 (#2)}
\newcommand{\PRC}[3]{{ Phys.~Rev. C} {\bf {#1}}, #3 (#2)}
\newcommand{\PRD}[3]{{ Phys.~Rev. D} {\bf {#1}}, #3 (#2)}
\newcommand{\PRL}[3]{{ Phys.~Rev.~Lett.} {\bf {#1}}, #3 (#2)}
\newcommand{\PR}[3]{{ Phys.~Rep.} {\bf {#1}}, #3 (#2)}
\newcommand{\ZPC}[3]{{ Z.~Phys. C} {\bf {#1}}, #3 (#2)}
\newcommand{\ZPA}[3]{{ Z.~Phys. A} {\bf {#1}}, #3 (#2)}
\newcommand{\JCP}[3]{{ J.~Comp.~Phys.} {\bf {#1}}, #3 (#2)}
\newcommand{\HIP}[3]{{ Heavy Ion Physics} {\bf {#1}}, #3 (#2)}

{}

\end{document}